\documentclass[12pt,showpacs,preprintnumbers,amsmath,amssymb]{revtex4}

\usepackage{times}
\usepackage{graphicx}
\usepackage{dcolumn}
\usepackage{bm}
\usepackage[usenames,dvipsnames]{color}
\usepackage{amsmath}
\usepackage[hypertex]{hyperref}
\usepackage{amsfonts}
\usepackage{amsthm}
\usepackage{amssymb}
\usepackage{mathrsfs}
\usepackage{subfigure}
\usepackage{ulem}
\usepackage{verbatim}
\usepackage{bbm}

\begin{document}

\title{Nonlocal correlations for manifold quantum systems:
Entanglement of two-spin states}

\author{K. Berrada$^{\textbf{1}}$} \email{kberrada@ictp.it}
\author{A. Mohammadzade$^{\textbf{2}}$}
\author{S. Abdel-Khalek$^{\textbf{1}}$}
\author{H. Eleuch$^{\textbf{3}}$}
\author{S. Salimi$^{\textbf{2}}$}
\affiliation{$^\textbf{1}$The Abdus Salam International Centre for Theoretical Physics, Strada Costiera 11, Miramare-Trieste, Italy\\
$^\textbf{2}$  Department of Physics, University of Kurdistan, P.O.Box 66177-15175 , Sanandaj, Iran\\
$^\textbf{3}$Institute for Quantum Studies and Department of Physics, \linebreak Texas A$\&$M
University, TX 77843, USA}

\begin{abstract}
In this paper, we study the bipartite entanglement of spin coherent states in the case of pure
and mixed states. By a proper choice of the subsystem spins, the entanglement for large class of quantum
systems is investigated. We generalize the result to the case of bipartite mixed states using a simplified expression of concurrence in Wootters' measure of the bipartite entanglement.  It is found that in some cases, the maximal entanglement of mixed states in the context of $su(2)$ algebra can be detected. Our observations may have important implications in exploiting these states in quantum information theory.
\end{abstract}

\maketitle

\section{Introduction}
$\indent$One of the most important issue in quantum theory is
quantum entanglement. This phenomenon has generated much interest
in the quantum information processing such as quantum computation
\cite {1,2}, quantum teleportation \cite {3}, quantum
cryptography \cite {4,5,6}, quantum error correction \cite {7} and more recently, metrology \cite{me}.
Therefore, it is interesting and important to quantify quantum
entanglement. The fundamental problem in quantum entanglement is to define the best measure
quantifying the amount of the entanglement of a given state. Some
measures such as concurrence \cite {13,14,15}, negativity \cite
{16,17,18,19}, tangle \cite {20,21} and linear entropy \cite {22}
can be used for quantifying entanglement \cite{aaa1,aaa2}. One of the best known
measures of entanglement is entanglement of formation  in a
bipartite systems, which is proposed by Bennett \textit{et. al} \cite {23}.
For a pure state in bipartite systems, this method is extensively
accepted. However, in the case of mixed states, it is heavily
dependent on the pure state decompositions and so far there is no
a general algorithm to find the minimum one. Another problem in
this phenomenon is to find a method to determine whether a given
state is entangled or not. One of the simplest states is
bipartite pure state. For example, $|\Psi_{AB}\rangle \in H_{AB}
= H_{A} \otimes H_{B}$ is called separable iff it can be written
as product of two vectors corresponding to Hilbert space of
subsystems: $|\Psi_{AB}\rangle = |\psi_{A}\rangle
|\phi_{B}\rangle$, otherwise it is entangled.

Another important concept widely used and applied in quantum information theory is the notion
of coherent or quasi-classical states. These states make a very useful tool for the investigation of
various problems in physics and have widespread applications in several fields of quantum physics
\cite{h,h1,h2,h3,h4}. Coherent states were originally constructed and developed for the
Heisenberg Weyl group to investigate quantized electromagnetic radiation \cite{gl} as the eigenstates
of the boson annihilation operator. Such states have the interesting property of minimizing the
Heisenberg uncertainty relation. Next, the following important coherent states are $SU(2)$ (spin) and
$SU(1,1)$ coherent states introduced by Perelomov \cite{pr1,pr2} which also have many applications
such as quantum optics, statistical mechanics and condensed matter physics \cite{ap1,ap2,hec1,hec2}.

Spin coherent states, also known as atomic coherent
states or angular momentum coherent states depending
on context, are analogous to the ordinary coherent states
of a harmonic oscillator in that they both may be considered
as pure, near-classical states of their corresponding systems. Recently, D. Markham and V. Vedral  have investigated the entanglement
as a result of the effect of a 50:50 beam splitter on spin coherent states using von Neumann entropy \cite{mar}. In the infinite spin limit, it is found that the spin coherent states are equivalent to the high-amplitude limit of the optical coherent states with zero entanglement in the output state. Furthermore, they have discussed an other aspects of classicality over the transition in the spin including the distinguishability using the representation of Majorana.  Nowadays, studying and understanding structures  of quantum entanglement using entangled non-orthogonal
states has received much attention; Bosonic  entangled coherent states
\cite {t1,t2,t3,t4} are the typical examples of entangled coherent states. From the point of view of
quantum  algebra applications to different physical systems, it is important to understand how
the  behavior of entangled coherent states is modified when the ordinary algebra is modified.  In this paper, we study the entangled spin coherent
states in the case pure and mixed states. In other words, we will consider the entanglement of
two-system spin states, of arbitrary spins $j_{1}$ and $j_{2}$, where each
system is prepared in spin coherent state  called entangled spin coherent states. In this process, we use
the concurrence as a measure and study its behavior for large class of quantum systems by an appropriate choice of the spins including qubit, qutrit, and qudit systems. The spin coherent states defined as  superpositions  of the angular momentum states which are nonclassical states regardless of the size of spins. In contrast, spin coherent states are classical-like
and become more so in the limit of large spin \cite{ge}.  Yet a bipartite entanglement
of spin coherent states, especially of distinguishable
spin coherent states, would be expected to have strong
nonclassical properties for different values of the spins.

This paper is organized as follows: In Sect. II we remind the basics of the $su(2)$ algebra and the associated coherent states, and we introduce the entangled spin coherent states.  In Sect. III using the concurrence as measure of the degree of entanglement, we study the
 entanglement behavior of these states in terms of different parameters for large class of quantum systems by a proper choice of the spins including qubit, qutrit, and qudit systems. In Sect. IV we generalize the results to the case of mixed states defined a statistical mixture of entangled spin coherent states  using a simplified expression of concurrence in Wootters' measure of the bipartite entanglement. Finally we summarize the paper in the Sect. V.

\section{Entangled spin coherent states}
Coherent states play a vital role in quantum physics,
particularly, in quantum optics \cite {24,25} and encoding quantum
information on continuous variables \cite {26,27}.  They also
have an important role in mathematical physics, for example, they
are very useful in performing stationary phase approximations to
path integral. One of the practical coherent state is $su(2)$
coherent state which can be  useful for describing   entangled coherent
states. The entangled coherent states are very
useful tools in different quantum transmission and processing tasks.

The spin coherent state can be obtained by applying successively the
raising operator on the state $|j,-j\rangle$
\begin{eqnarray}
\nonumber|Z,j\rangle&=& R(Z)|j,-j\rangle=\exp\left[-\frac{1}{2}\left(J_{+}e^{-i\varphi}-J_{-}e^{i\varphi}\right)\right]|j,-j\rangle\\
 &=& \frac{1}{(1+\mid Z \mid^{2})^{j}}
 \sum^{j}_{m=-j}\left[\frac{(2j)!}{(j+m)!(j-m)!}\right]^{\frac{1}{2}}Z^{j+m}|j,m\rangle,
\end{eqnarray}
where $R(Z)$ is the rotation operator, $J_{-}$ and $J_{+}$ are
the lowering and raising operators of the $su(2)$ Lie algebra,
respectively. The generators of the $su(2)$ Lie algebra, $J_{\pm}$
and $J_{z}$, satisfy the following commutation relations
\begin{equation}
[J_{+}, J_{-}] = 2J_{z} \qquad [J_{z},J_{\pm}] = \pm J_{\pm}.
\end{equation}
These generators act on an irreducible unitary representation as
follows
\begin{equation}J_{\pm}|j,m\rangle = \sqrt{(j \mp m)(j \pm m + 1)}|j,m \pm
1\rangle;\qquad
 J_{z}|j,m\rangle = m|j,m\rangle.
\end{equation}
Here, we are going to investigate the pure states of the following
form
\begin{equation}\label{en}
|\psi\rangle =
N\left[|Z_{1},j_{1}\rangle\otimes|-Z_{2},j_{2}\rangle+e^{i\phi}|-Z_{1},j_{1}\rangle\otimes|Z_{2},j_{2}\rangle\right],
\end{equation}
where $|Z_{1},j_{1}\rangle$ and $|-Z_{1},j_{1}\rangle$ are
normalized states of the spin 1 and $|Z_{2},j_{2}\rangle$ and
$|-Z_{2},j_{2}\rangle$ are states of the spin 2, such as the inner
product for subsystems are as following:
\begin{equation}
P_{1}=\langle Z_{1},j_{1}|-Z_{1},j_{1} \rangle =
\frac{\left(1- \mid Z_{1}\mid^{2}\right)^{2j_{1}}}{\left(1+ \mid Z_{1} \mid^{2}\right)^{2j_{1}}}
\end{equation}

\begin{equation}
P_{2}=\langle Z_{2},j_{2}|-Z_{2},j_{2} \rangle =
\frac{\left(1- \mid Z_{2}
\mid^{2}\right)^{2j_{2}}}{\left(1+ \mid Z_{2} \mid^{2}\right)^{2j_{2}}}.
\end{equation}
The normalization factor $N$ is
\begin{equation}
N={1\over\sqrt{2}}\left[1+\cos\phi\left({\left(1-|Z_1|^2\right)\over\left(1+|Z_1|^2\right)}\right)^{2j_1}
\left({\left(1-|Z_2|^2\right)\over\left(1+|Z_2|^2\right)}\right)^{2j_2}\right]^{-{1\over2}}.
\end{equation}

We can show that the bipartite state (\ref{en}) can be considered as a two-qubit system by introducing
an orthogonal normalized basis in the subspace spanned by $|Z_{1},j_{1}
\rangle$ and $|-Z_{1},j_{1}\rangle$ and by $|Z_{2},j_{2}\rangle$ and $|-Z_{2},j_{2}\rangle$ as \cite{en1,en2}

\begin{equation}
|0\rangle =|Z_{1},j_{1} \rangle\qquad |1\rangle=\frac{|-Z_{1},j_{1}
\rangle -P_{1}|Z_{1},j_{1} \rangle }{N_{1}}
\end{equation}
or in other words
$$
|Z_{1},j_{1}\rangle=|0\rangle \qquad|-Z_{1},j_{1}\rangle=N_{1}|1\rangle
+P_{1}|0\rangle\quad \;\mathrm{for}\; \mathrm{spin} \;1,
$$
\begin{equation}
|0\rangle =|Z_{2},j_{2} \rangle \qquad |1\rangle=\frac{|-Z_{2},j_{2}
\rangle -P_{2}|Z_{2},j_{2} \rangle }{N_{2}}
\end{equation}
or in other words
\begin{equation}
|Z_{2},j_{2}\rangle=|0\rangle\qquad |-Z_{2},j_{2}\rangle=N_{2}|1\rangle
+P_{2}|0\rangle\quad \;\mathrm{for}\; \mathrm{spin} \;2
\end{equation}
where
\begin{equation}
N_{1}=\left[1-|P_1|^2\right]^{1\over2}\qquad \mathrm{and} \qquad N_{2}=\left[1-|P_2|^2\right]^{1\over2}.
\end{equation}
In this basis, the bipartite entangled spin coherent state can be written as
\begin{equation}\label{rep}
|\psi\rangle = \left(NP_{2}+Ne^{i\phi}P_{1}\right)|00\rangle +NN_{2}|01\rangle
+ Ne^{i\phi}N_{1}|10\rangle.
\end{equation}
This state can describe the nonlocal correlations for a large class of bipartite states including qubit, qutrit, and qudit systems through an appropriate choice of the spins $j_1$ and $j_2$. The entangled spin coherent states can be experimentally prepared by analogy with the generation of the entangled coherent states for harmonic oscillators \cite{pre}.

\section{Concurrence for different bipartite states in the context of $su(2)$ algebra}
In this section, we derive the amount of entanglement of two-spin system states and
study its behavior  in terms of the parameters involved in the coherent states.
Here, we adopt the concurrence to characterize and quantify the degree of entanglement of the
bipartite state via entangled spin coherent states.

For quantifying the degree of entanglement of the bipartite spin states, we consider the state (\ref{en}) in the framework of 2$\times$2 Hilbert space as mentioned in the above section. In this considered case, the concurrence of the entangled spin coherent states takes the form
\begin{equation}
C= 2N^{2}\left[\left(1-|P_1|^2\right)\left(1-|P_2|^2\right)\right]^{1\over2}.
\end{equation}
Generally, a pure state  is referred as a separable state when $C=0$ and it maximally entangled state for
$C=1$.

Let us investigate the degree of entanglement of the two-spin state in terms of different parameters that specify the coherent states. From the inner product of each subsystem, we
can see that for the case where $|Z_1|=|Z_2|=1$, i.e., $Z_1=\pm e^{i\theta_1}$ and $Z_2=\pm e^{i\theta_2}$, the states $\langle\pm e^{i\theta_1},j_1|\mp e^{i\theta_1},j_1\rangle=\langle \pm e^{i\theta_2},j_2|\mp e^{i\theta_2},j_2\rangle=0$ are orthogonal, with the corresponding spin coherent
states
\begin{equation}
|\pm e^{i\theta_k},j_k\rangle={1\over 2^{j_k}}\sum_{m_k=-j_k}^{j_k}\left[{(2j_k)!\over(j_k+m_k)!(j_k-m_k)!}\right]^{1\over2}\left({\pm e}^{i(j_k+m_k)\theta_k}\right)|j_k,m_k\rangle;\qquad k=1,2.
\end{equation}
In this considered case, we obtain  maximal entanglement ($C=1$) of the  entangled spin coherent state (\ref{en}) as long as $Z_1=\pm e^{i\theta_1}$ and $Z_2=\pm e^{i\theta_2}$ for all nonzero values of the spins ($j_1$, $j_2$) and the relative phase $\phi$. Then from equation (\ref{en}), we have a set of mutually orthogonal Bell states of the form

\begin{equation}
|\psi\rangle={1\over\sqrt{2}}\left[| \pm e^{i\theta_1},j_1\rangle\otimes|\mp e^{i\theta_2},j_2\rangle+e^{i\phi}|\mp e^{i\theta_1},j_1\rangle\otimes|\pm e^{i\theta_2},j_2\rangle\right].
\end{equation}

In order to observe the influence of the parameters that specify the coherent states on the
 entanglement behavior  of the bipartite system state (\ref{en}), the concurrence of the two-spin state is
plotted in figures $(1)$ and $(2)$ in terms of the both amplitudes for various values  of spins. Such a state exhibits several quantum systems with different dimension  by an appropriate choice of the spins $j_1$ and $j_2$.  From figures $(1)$ and $(2)$, we can see evident differences of the entanglement for the different quantum systems. In fact, when the condition $Z_1=\pm e^{i\theta_1}$ and $Z_2=\pm e^{i\theta_2}$ does not hold, we find  strong entanglement (concurrence tends to approach maximal value) as long as one of the amplitudes is near $\pm e^{i\theta_k}$ while the other differs significantly from $\pm e^{i\theta_k}$. As the both amplitudes differ significantly from $\pm e^{i\theta_k}$, the concurrence becomes weaker and nonexistent for large enough deviation from $\pm e^{i\theta_k}$. On other hand, the amount of  entanglement increases, is shown as  the  spins $j_1$ and $j_2$ increase. From these results, it is clear that the amplitude values  can restrain the entanglement of the two-spin system as they get far from $\pm e^{i\theta_k}$. However, the spins  $j_1$ and $j_2$ can enhance the entanglement of the system.

\newpage
\begin{center}
 \includegraphics[width=8.5cm]{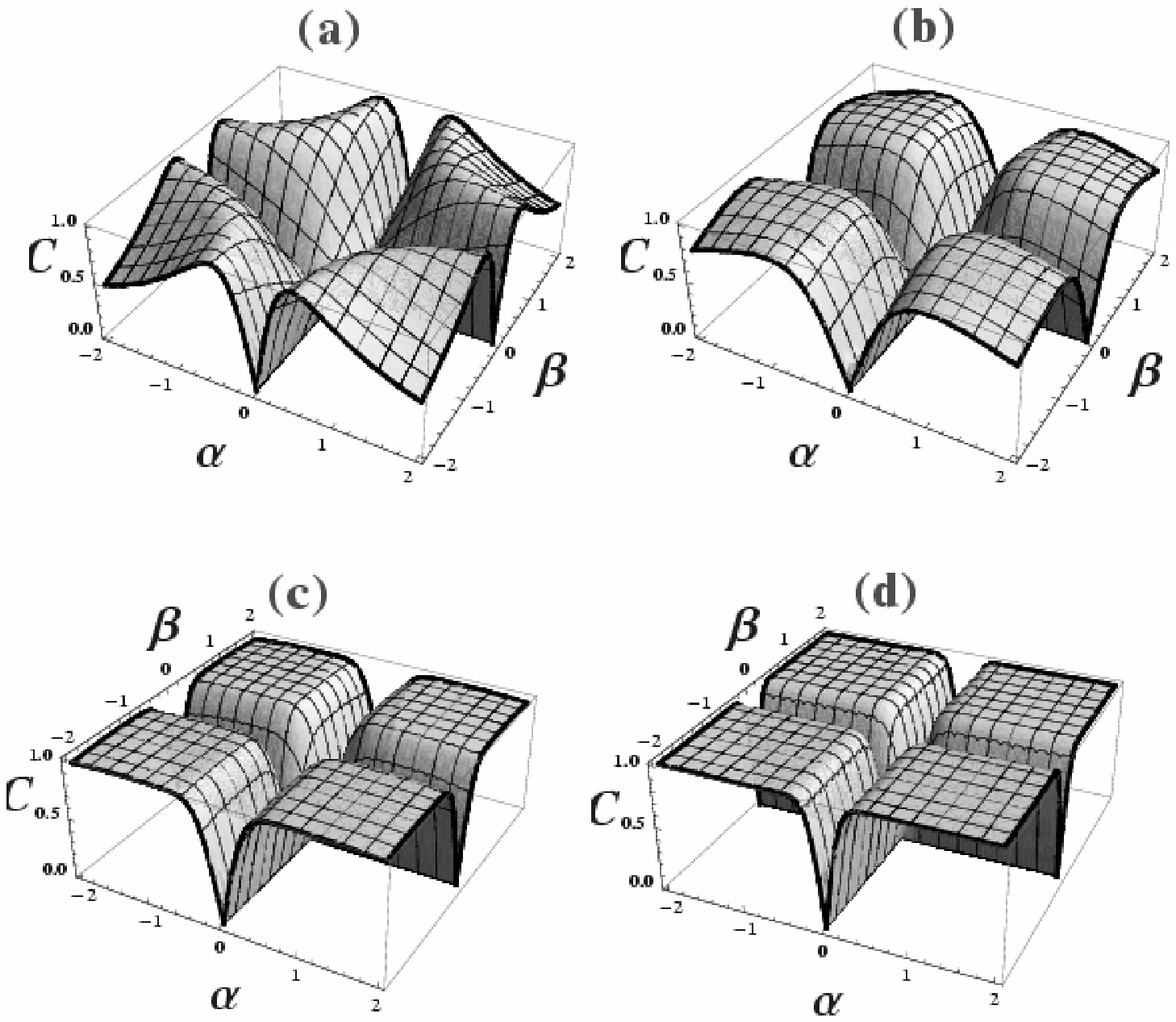}\\

 \underline{Figure 1. Entanglement as a function of the coherent state amplitudes for $\phi=0$.}
 \underline{ ($a$) qubit-qubit system ($j_1=j_2=0.5$), ($b$) qutrit-qutrit system ($j_1=j_2=1$)}
 \underline{ ($c$) qudit-qudit system ($j_1=j_2=2$), and ($d$) qudit-qudit system ($j_1=j_2=4$)}
 \end{center}

 \begin{center}
 \includegraphics[width=8.5cm]{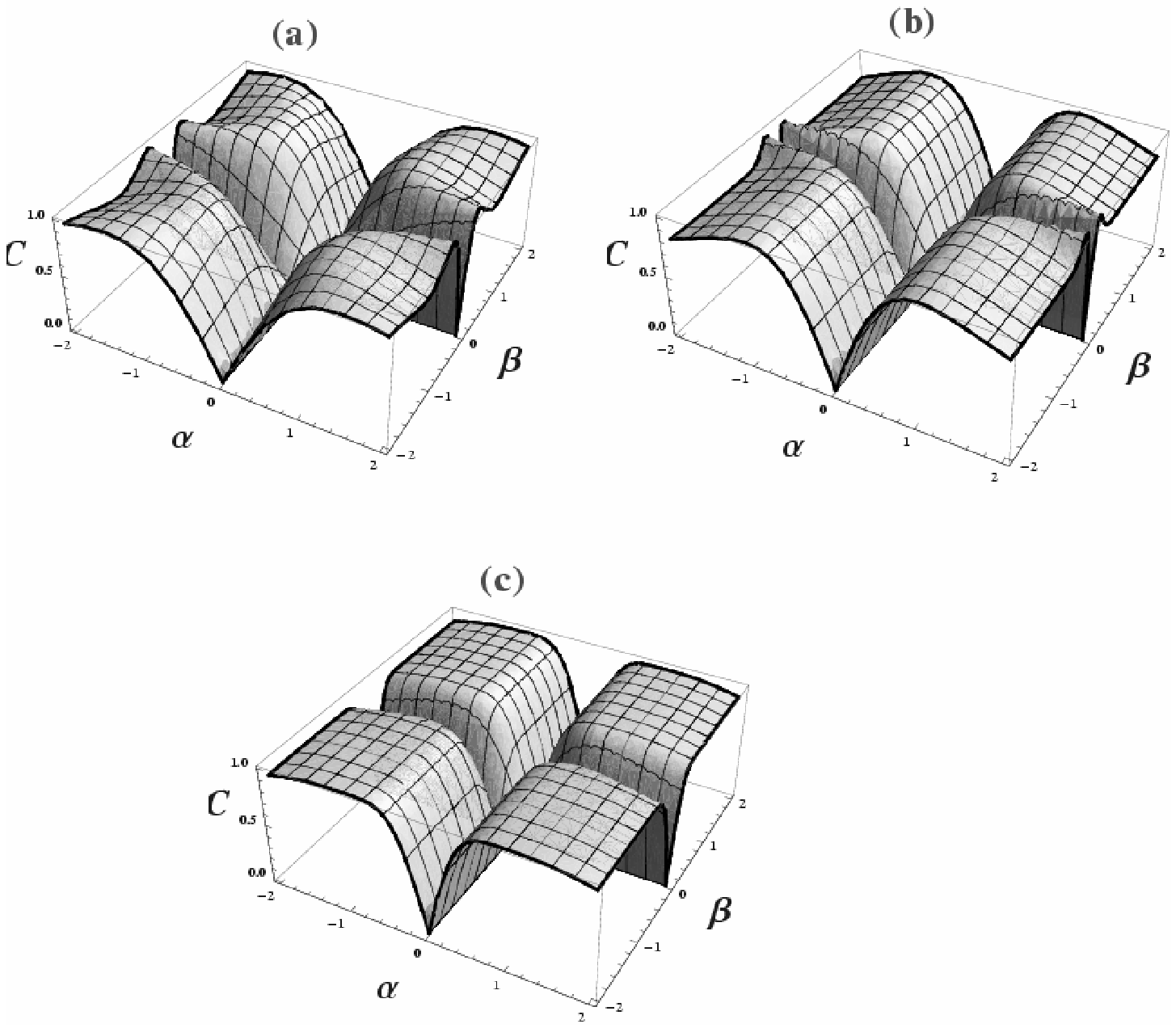}\\

 \underline{Figure 2. Entanglement as a function of the coherent state amplitudes for $\phi=0$.}
 \underline{ ($a$) qubit-qutrit system ($j_1=0.5 \;;\; j_2=1$), ($b$) qubit-qudit system ($j_1=0.5 \;;\; j_2=4$),}
 \underline{ and ($c$) qutrit-qudit system ($j_1=1 \;;\; j_2=4$).}
 \end{center}

\section{The case of mixed states}

In the case of mixed states, the bipartite system quantum state must be represented not by
a bracket as the case of pure states, but by a matrix called density matrix and denoted by $\rho$ in
quantum mechanics. It is always to decompose the density matrix $\rho$ into a classical mixture of the
density matrices of a set of pure states $|\psi_i\rangle$. In the proposed mixed state, we have
\begin{equation}\label{mix}
\rho = \sum_{i}p_{i}|\psi_{i}\rangle\langle\psi_{i}|,
\end{equation}
where $\{|\psi_i\rangle\}$ are distinct normalized  pure states of the bipartite system given by
\begin{equation}
|\psi_i\rangle =
N^i\left[|Z^i_{1},j_{1}\rangle\otimes|-Z^i_{2},j_{2}\rangle+e^{i\phi}|-Z^i_{1},j_{1}\rangle\otimes|Z^i_{2},j_{2}\rangle\right]
\end{equation}
and $\{p_i\}$ are the corresponding probabilities (i.e., $p_i\geq0$
and $\sum_ip_i=1$).
\\There is a condition for
separability and inseparability of mixed states like the pure
states which are mentioned above. A mixed state $\rho$ via entangled spin coherent states is said to
be separable  if it can be written as a convex sum of separable
pure states, i.e., $\rho =
\sum_{i}p_{i}\rho_{i}^{(j_1)}\otimes\rho_{i}^{(j_2)}$, where
$\rho_{i}^{(j_1,j_2)}$ is the reduced density operator of subsystem
($j_1$,$j_2$), respectively, given by
$\rho_{i}^{(j_1,j_2)}=Tr_{(j_2,j_1)}\left(|\psi_{i}\rangle\langle\psi_{i}|\right)$.
The state $\rho$ is entangled if it cannot be represented as a
mixture of a separable pure states.

One of the difficulties in the quantification and characterization of mixed entanglement is linked
to the fact that the entanglement of a superposition of pure bipartite states cannot be simply
expressed as a function of the entanglement of the individual states in the superposition. This is
because entanglement mostly depends on the coherence among the states in the superposition. It is
therefore somewhat surprising that there exist tight lower and upper bounds on the entanglement of
a superposition of states in terms of the entanglement of the individual states in the superposition.
Here, we shall use the concept of the concurrence for quantifying the amount of entanglement of
bipartite system mixed states in the context of the $su(2)$ algebra, by introducing a
simplified expression of the concurrence in Wootters's measure of entanglement of bipartite system
states. Indeed, using the same technique as mentioned previously, one can consider the state (\ref{mix}) as a two-qubit system mixed state and we write it in the standard basis, $\{|00\rangle, |01\rangle, |10\rangle, |11\rangle\}$, as
\begin{equation}
\rho=\sum_ip_i|\psi_i\rangle\langle\psi_i|,
\end{equation}
where $\{|\psi_i\rangle\}$ are the pure states of bipartite system defined as
\begin{equation}\label{eng}
|\psi_i\rangle=\left(N^iP_2^i+N^ie^{i\phi_i}P^i_1\right)|00\rangle+N^iN_2^i|01\rangle+N^ie^{i\phi_i}N_1^i|10\rangle.
\end{equation}
with
\begin{eqnarray}
\nonumber N^i&=&{1\over\sqrt{2}}\left[1+\cos\phi_i\left({\left(1-|Z^i_1|^2\right)\over\left(1+|Z^i_1|^2\right)}\right)^{2j_1}
\left({\left(1-|Z^i_2|^2\right)\over\left(1+|Z^i_2|^2\right)}\right)^{2j_2}\right]^{-{1\over2}}\\
\nonumber N^i_{1}&=&\left[1-|P^i_1|^2\right]^{1\over2}\\
 N^i_{2}&=&\left[1-|P^i_2|^2\right]^{1\over2}
\end{eqnarray}
and
\begin{eqnarray}
\nonumber P^i_{1}&=&\frac{\left(1- \mid Z^i_{1}\mid^{2}\right)^{2j_{1}}}{\left(1+ \mid Z^i_{1} \mid^{2}\right)^{2j_{1}}}\\
P^i_{2}&=&\frac{\left(1- \mid Z^i_{2}
\mid^{2}\right)^{2j_{2}}}{\left(1+ \mid Z^i_{2} \mid^{2}\right)^{2j_{2}}}.
\end{eqnarray}

We can define the concurrence of the mixed state $\rho$ as a
convex
 roof method which is the average concurrence of an ensemble pure states
 of the decomposition, minimized over all decomposition of
 $\rho$ \cite {31}
\begin{equation}\label{4}
C(\rho) = \inf\sum_{i}p_{i}C(|\psi_{i}\rangle),
\end{equation}
where $C(|\psi_{i}\rangle)$ is the concurrence of the pure state
$|\psi_{i}\rangle$ given by (\ref{eng}). Wootters and Hill have found
an explicit formula of the concurrence defined as \cite {32}
\begin{equation}
C(\rho) = \max\{\lambda_{1} - \lambda_{2} - \lambda_{3} -
\lambda_{4},0\}.
\end{equation}
Here $\lambda_{i}$ is the square root of eigenvalues of
$\rho(\sigma_{y}\otimes\sigma_{y})\rho^{\ast}(\sigma_{y}\otimes\sigma_{y})$
in decreasing order ($\rho^{\ast}$ denotes the complex conjugate of
$\rho$).
\\In general, for a bipartite system mixed  state with no more than
two-non-zero eigenvalues ($\mu_1,\mu_2$), there is an
explicit formula of the square of the concurrence defined as \cite{ka,ka1,ka2}
\begin{equation}\label{con}
C^{2}(\rho) = \left(\mu_{1}^{2}C_{1}^{2} + \mu_{2}^{2}C_{2}^{2}
 \right)
  + \frac{1}{2}\mu_{1}\mu_{2}\left| \textbf{c}_{+}-\textbf{c}_{-} \right|^{2}
  - \frac{1}{2}\mu_{1}\mu_{2}\left| (\textbf{c}_{+}-\textbf{c}_{-})^{2} - 4\textbf{c}_{1}\textbf{c}_{2}\right|,
 \end{equation}
where

  \begin{equation}
 C_{i} =  \mid \textbf{c}_{i}\mid  =  2\mid a_{i}d_{i}-b_{i}c_{i}\mid \end{equation}
is the concurrence of the pure state $|\mu_{i}\rangle$, and

 \begin{equation}C_{\pm} = \mid \textbf{c}_{\pm}\mid =
 \frac{1}{2}\left|(a_{1} \pm a_{2})(d_{1} \pm d_{2}) - (b_{1} \pm b_{2})(c_{1} \pm c_{2})\right|
 \end{equation}
  is the concurrence of the pure state
  $|\mu_{\pm}\rangle =
  \frac{1}{\sqrt{2}}(|\mu_{1}\rangle\rangle \pm |\mu_{2}\rangle).$ $\textbf{c}_{i}$
and $\textbf{c}_{\pm}$ are the corresponding complex concurrences. Here the pure states $|\mu_1\rangle$ and $|\mu_2\rangle$ are the eigenvectors
of the mixed state  given by
\begin{eqnarray}
\nonumber|\mu_1\rangle&=&a_1|00\rangle+b_1|01\rangle+c_1|10\rangle+d_1|11\rangle\\
|\mu_1\rangle&=&a_2|00\rangle+b_2|01\rangle+c_2|10\rangle+d_2|11\rangle.
\end{eqnarray}

The advantages of the concurrence given by Eq. (\ref{con}) are expressed the concurrence of the
mixed state as a function of the concurrence of the pure states and their simple combinations, also
it can be solved easily analytically. Furthermore, it reveals some general features and exhibits of
important results in quantum information area.

Let us consider a class of mixed states given by a statistical mixture of two bipartite pure via entangled spin coherent states
\begin{eqnarray}
\nonumber\rho&=&\sum_{i=1,2}p_i|\psi_i\rangle\langle\psi_i|\\
&=&p_1|\psi_1\rangle\langle\psi_1|+p_1|\psi_2\rangle\langle\psi_2|
\end{eqnarray}
where, the pure states $|\psi_{1}\rangle$ and $|\psi_{2}\rangle$ are defined as
{\small\begin{eqnarray}
\nonumber|\psi_{1}\rangle &=&
N^1[|Z^1_{1},j_{1}\rangle|-Z^1_{2},j_{2}\rangle+e^{i\phi_1}|-Z^1_{1},j_{1}\rangle|Z^1_{2},j_{2}\rangle]
\equiv (N^1P^1_{2}+N^1e^{i\phi_1}P^1_{1})|00\rangle +N^1N^1_{2}|01\rangle +
N^1e^{i\phi_1}N^1_{1}|10\rangle\\
\nonumber|\psi_{2}\rangle&=&
N^{2}[|Z_{1}^{2},j_{1}\rangle|-Z_{2}^{2},j_{2}\rangle+e^{i\phi_2}|-Z_{1}^{2},j_{1}^{2}\rangle|Z_{2}^{2},j_{2}\rangle]
\equiv (N^{2}P_{2}^{2}+N^{2}e^{i\phi_2}P_{1}^{2})|00\rangle
+N^{2}N_{2}^{2}|01\rangle + N^{2}e^{i\phi_2}N_{1}^{2}|10\rangle.\\
\end{eqnarray}}
For such states, $C_1$ and $C_2$
are given respectively by
\begin{equation}
C_1=\left| \textbf{c}_{1} \right|=2\left(N^1\right)^{2}N^1_{1}N^1_{2}
\end{equation}
\begin{equation}
C_2=\mid \textbf{c}_{2}
\mid=2\left(N^2\right)^{2}N_{1}^{2}N_{2}^{2}.
\end{equation}
Therefore, the concurrence of the mixed state is
\begin{equation}
C^{2}(\rho) = \left(p_{1}^{2}C_{1}^{2} +
p_{2}^{2}C_{2}^{2}\right)
  + \frac{1}{2}p_{1}p_{2}\mid \textbf{c}_{+}-\textbf{c}_{-} \mid^{2}
  - \frac{1}{2}p_{1}p_{2}\mid (\textbf{c}_{+}-\textbf{c}_{-})^{2} - 4\textbf{c}_{1}\textbf{c}_{2}\mid
\end{equation}
where
\begin{equation}
C_{\pm}=\mid\textbf{c}_{\pm}\mid= \left|(N^1N^1_{2}\pm
N^{2}N_{2}^{2})(N^1N^1_{1}e^{i\phi_1} \pm N^{2}N_{1}^{2}e^{i\phi_2})
\right|
\end{equation}
is the concurrence of the pure state
\begin{equation}
|\psi_{\pm} \rangle =\frac{1}{\sqrt{2}}(|\psi_{1} \rangle \pm
|\psi_{2} \rangle).
\end{equation}
Many works about quantifying and characterizing bipartite mixed state entanglement have
been proposed using several schemes \cite{sc1,sc2}. Basing on the simplified expression (\ref{con}), we may
present observable lower and upper bounds of the squared concurrence for the bipartite states in
the context of $su(2)$ algebra,
\begin{equation}
 (p_{1}C_{1} - p_{2}C_{2})^{2}\leq C^{2}(\rho)
 \leq(p_{1}C_{1} + p_{2}C_{2})^{2},
 \end{equation}
where
\begin{equation}
(p_{1}C_{1} - p_{2}C_{2})^{2} =4\left(p_{1}
\left(N^1\right)^{2}N^1_{1}N^1_{2} -p_{2}
\left(N^2\right)^{2}N_{1}^{2}N_{2}^{2} \right)^{2}
\end{equation}
and
\begin{equation}
(p_{1}C_{1} + p_{2}C_{2})^{2} =4\left(p_{1}
\left(N^1\right)^{2}N^1_{1}N^1_{2} +p_{2}
\left(N^2\right)^{2}N_{1}^{2}N_{2}^{2} \right)^{2},
\end{equation}
are respectively the lower and upper bounds of concurrence. At this point the above inequality can prove some important features by a proper
choice of the parameters that specify the spin coherent states:\\
-When $(\textbf{c}_{+}-\textbf{c}_{-})^2\geq
4\textbf{c}_{1}\textbf{c}_{2}\geq0$, the concurrence via entangled spin coherent states is equal to the
upper bound,
\begin{equation}
C^{2}(\rho) =4\left(p_{1}
\left(N^1\right)^{2}N^1_{1}N^1_{2} +p_{2}
\left(N^2\right)^{2}N_{1}^{2}N_{2}^{2} \right)^{2}.
\end{equation}
-For $0\leq \left(\textbf{c}_{+}-\textbf{c}_{-}\right)^2\leq
4\textbf{c}_{1}\textbf{c}_{2}$, the square of concurrence  becomes,
\begin{equation}
C^{2}(\rho) =4\left[\left(p_{1}\left(N^1\right)^{2}N^1_{1}N^1_{2}
-p_{2}\left(N^2\right)^{2}N_{1}^{2}N_{2}^{2}\right)^{2}
+p_{1}p_{2}\left|N^1N^1_{2}N^{2}N^{2}_{1}e^{i\phi_2}+N^{2}N_{2}^{2}N^1N^1_{1}e^{i\phi_1}\right|^{2}\right].
\end{equation}
-If $ \textbf{c}_{1}\textbf{c}_{2}\leq 0$, the concurrence is equal
to the lower bound,
\begin{equation}
C^{2}(\rho)=4\left(p_{1}
\left(N^1\right)^{2}N^1_{1}N^1_{2} -p_{2}
\left(N^2\right)^{2}N_{1}^{2}N_{2}^{2} \right)^{2}
\end{equation}
-For $\textbf{c}_{+}=\textbf{c}_{-} $, we have
\begin{equation}
N^1_{2}N^{2}_{1}e^{i\phi_2}+N^2_{2}N^1_{1}e^{i\phi_1}=0
\end{equation}
and the concurrence reaches the lower bound,
\begin{equation}
C^{2}(\rho)=4\left(p_{1}
\left(N^1\right)^{2}N^1_{1}N^1_{2} -p_{2}
\left(N^2\right)^{2}N_{1}^{2}N_{2}^{2} \right)^{2}.
\end{equation}

We now study the behavior of the concurrence of bipartite system mixed state via entangled spin coherent states in terms of the
coherent state parameters $P_i, P_i^{'}$ and the probabilities $p_i$  according to the bipartite pure state conditions
which are previously discussed. To see the effects of different parameters on the square of the concurrence of the mixed state, we consider the case where either of pure states is maximally entangled.  In figures ($3$) and ($4$), by choosing $Z_1^{2}=Z_2^{2}=1$, we display the variation of $C^2\left(\rho\right)$  as a function of the amplitudes $Z^1_1$ and $Z^1_2$ for $p_1=p_2=1/2$.  From the figures we can see a direct monotonic relationship between  the mixed state entanglement and its pure states one in the context of $su(2)$ algebra,  exhibiting the same behavior for different ranges of the amplitudes in the different quantum systems. In this way, the mixed states defined as a statistical mixture of entangled spin coherent states may be very particular, exhibiting an entanglement  behavior like pure states. In this case we have two important cases: When the amplitudes of pure states equal $\pm e^{i\theta_k}$, the mixed state concurrence tends to its maximal value $C(\rho)=1$ (see figure 3). As the amplitudes get far from the condition $\pm e^{i\theta_k}$, the maximal value of entanglement of the mixed state decreases, and the concurrence is bounded $0<C(\rho)<1$ (see figure 4). From these results, we find that the pure state amplitudes of the mixed state in the framework of the $su(2)$ algebra may damage the amount of entanglement.

\begin{center}
 \includegraphics[width=8.5cm]{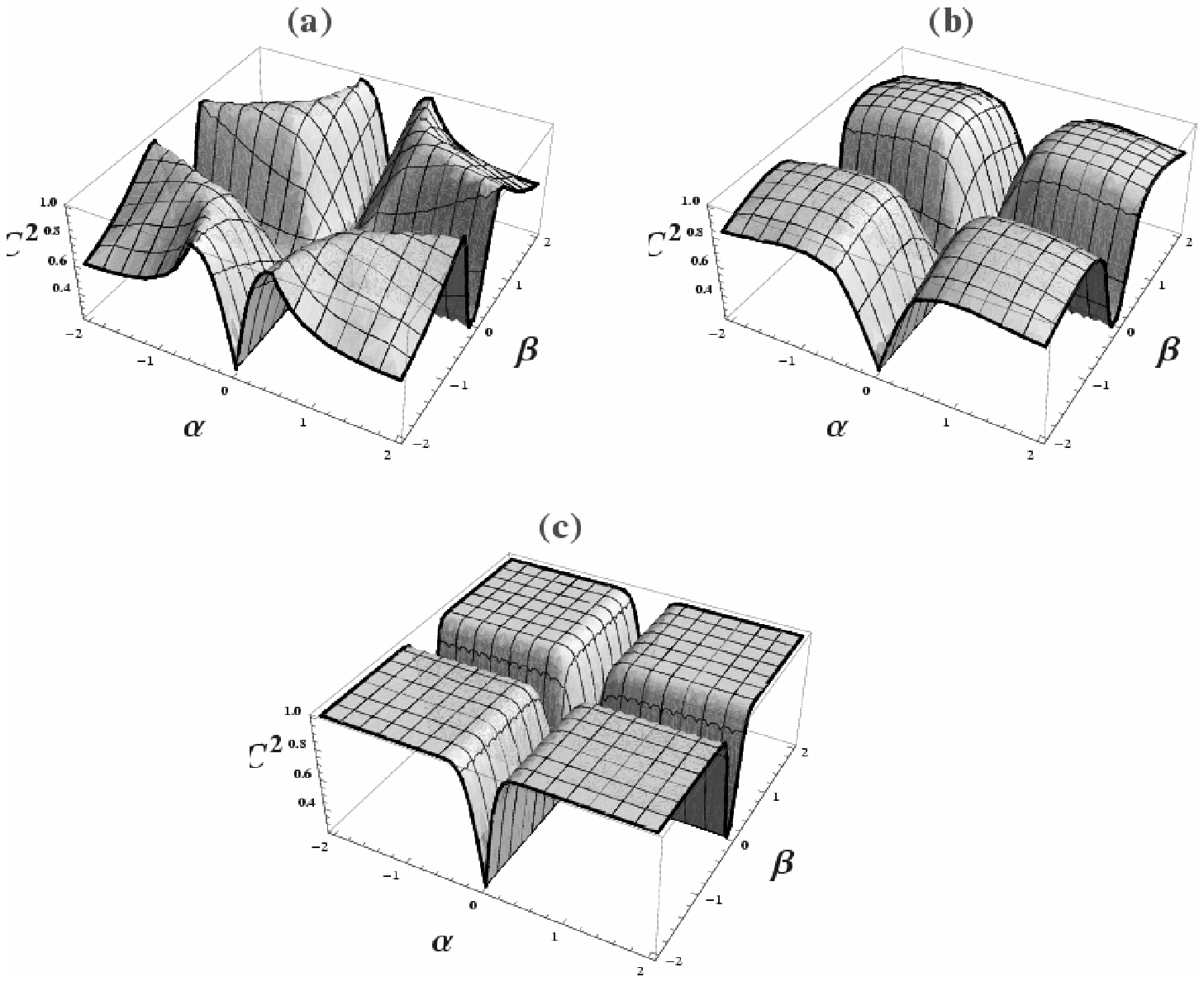}\\

 \underline{Figure 3. Entanglement as a function of the coherent state amplitudes for $Z_1^{2}=Z_2^{2}=1$ and $p_1=p_2=0.5$.}
 \underline{ ($a$) qubit-qubit system ($j_1= j_2=0.5$), ($b$) qutrit-qutrit system ($j_1= j_2=1$),}
 \underline{ and ($c$) qudit-qudit system ($j_1= j_2=4$). }
 \end{center}

\begin{center}
 \includegraphics[width=8.5cm]{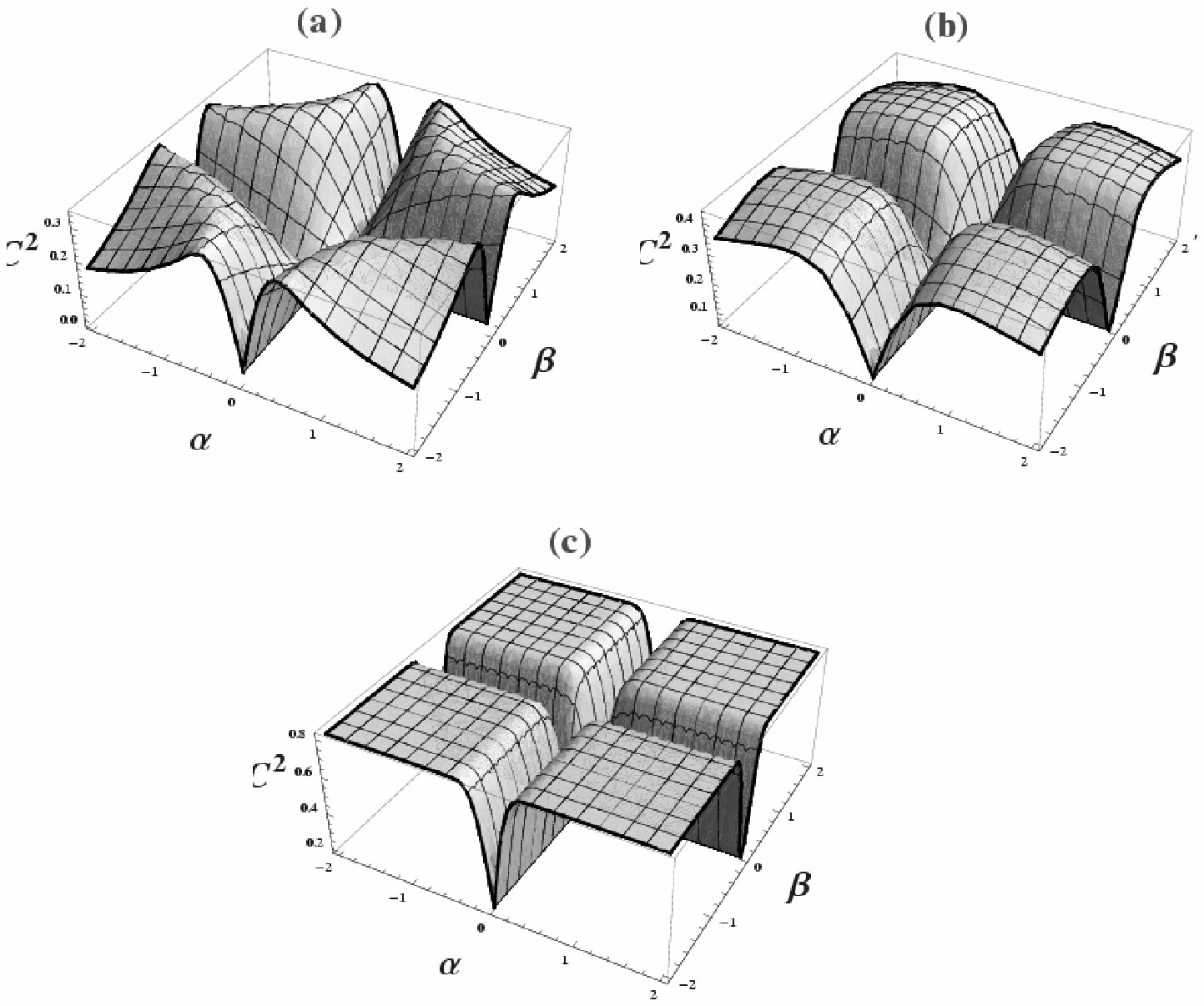}\\

 \underline{Figure 4. Entanglement as a function of the coherent state amplitudes for $Z_1^{2}=Z_2^{2}=4$ and $p_1=p_2=0.5$.}
 \underline{ ($a$) qubit-qubit system ($j_1= j_2=0.5$), ($b$) qutrit-qutrit system ($j_1= j_2=1$),}
 \underline{and ($c$) qudit-qudit system ($j_1= j_2=4$).}
 \end{center}

\section{Summary}
In this paper, we investigated the bipartite entanglement in the framework of $su(2)$ quantum algebra via entangled spin coherent states in the case of pure and mixed states including a large class of quantum systems by an appropriate choice of the subsystem spins and consider a possible application
in different contexts of quantum physics.

For two-spin pure states, using a change of basis, we have investigated the behavior of entanglement in the context of $su(2)$ algebra in terms
of the parameters that specify the spin coherent states. We have obtained the condition under which the entangled spin coherent states become maximally entangled for different systems. According to the condition, a large class of Bell states are found for any choice of the subsystem spins. In the case of the mixed states, using a simplified expression of the concurrence in terms
of the concurrences of the pure states and their simple combinations, we have calculated  and written the upper and lower bounds
of the concurrence  and studied its behavior as a function of the amplitudes of
coherent states and the probabilities for a class of mixed states.  In this considered case, the maximal entanglement of the mixed state of two-spin states can be detected.

It  will be important to  study the  multipartite entanglement in the context of the $su(2)$ algebra  which make a useful
contribution for more understanding the entanglement behavior  in these systems, and consider the possible applications in the various quantum
information processing and transmission tasks.

\section*{ACKNOWLEDGMENTS} K. B.  would like to thank the Abdus Salam centre for Theoretical Physics
for helpful hospitality. H. E. pleased to express his sincere gratitude for the hospitality at the Max Planck
Institute for the Physics of Complex Systems in Dresden.

\newpage

\end{document}